\documentclass[aip,jcp,twocolumn,superscriptaddress,reprint]{revtex4-1}
\usepackage{graphicx, color}
\usepackage{amsmath, amssymb, amsfonts, mathrsfs,ulem,comment,subfigure,ulem}
\usepackage{times}
\usepackage[margin=1in]{geometry}
\usepackage{graphicx}
\usepackage{dcolumn}
\usepackage{bm}

\usepackage[utf8]{inputenc}
\usepackage[T1]{fontenc}
\usepackage{mathptmx}

\begin{document}

\title{Ionic Structure and Decay Length in Highly-Concentrated Confined Electrolytes}
\author{Nasim Anousheh}
\affiliation{Intelligent Systems Engineering, 700 N. Woodlawn Avenue, Indiana University, Bloomington, Indiana 47408}
\author{Francisco J. Solis}
\affiliation{School of Mathematical and Natural Sciences, Arizona State University, Glendale, Arizona,  85306}
\author{Vikram Jadhao}
\email{vjadhao@iu.edu}
\affiliation{Intelligent Systems Engineering, 700 N. Woodlawn Avenue, Indiana University, Bloomington, Indiana 47408}


\begin{abstract}
We use molecular dynamics simulations of the primitive model of electrolytes to study the ionic structure in aqueous monovalent electrolyte solutions confined by charged planar interfaces over a wide range of electrolyte concentration, interfacial separation, surface charge density, and ion size. The investigations are inspired by recent experiments that have directly measured the increase in the decay length for highly-concentrated electrolytes with increase in concentration. The behavior of ions in the nanoconfinement created by the interfaces is probed by evaluating the ionic density profiles, net charge densities, screening factors, and decay length associated with the screening of the charged interface. Results show the presence of two distinct regimes of screening behavior as the concentration is changed from 0.1 M to 2.5 M for a wide range of electrolyte systems generated by tuning the interfacial separation, surface charge density, and ionic size. For low concentrations, the screening factor exhibits a monotonic decay to 0 with a decay length that decreases sharply with increasing concentration. For high concentrations ($\gtrsim 1$ M), the screening factor has a non-monotonic behavior signaling charge inversion and formation of structured layers of ions near the interfaces. The decay length under these conditions rises with increasing concentration, exhibiting a power-law behavior. To complement the simulation results, a variational approach is developed that produces charge densities with characteristics consistent with those observed in simulations. The results demonstrate the relation between the rise in the strength of steric correlations and the changes in the screening behavior.
\end{abstract}

\maketitle

\section{Introduction}

The behavior of electrolyte ions in liquids confined between charged macromolecules regulate many processes in soft materials such as colloids, emulsions, polymeric membranes, and proteins \cite{levin1,honig}. The local ionic environment can modulate the effective interaction between the charged macromolecules, thus changing their assembly behavior \cite{linse1999electrostatic,jos4}. Many energy storage applications \cite{cayre2007polyelectrolyte,siwy2002fabrication,he2009tuning,perry2010ion,zhang2018bioinspired} and separation process technologies \cite{faucher2019critical,park2017maximizing,werber2016materials} based on these materials rely on the complex organization and transport of ions near macromolecular surfaces.  
Therefore, investigating the self-assembly of ions in nanoconfinement created by material surfaces has been the focus of many experimental and theoretical studies \cite{schlossman1,laanait2012tuning,allen1,boda,dos2015electrolytes,jing2015ionic,podgornik1,zwanikken1,wang1,kjellander1985inhomogeneous,feng2010ion,fahrenberger2014simulation,qiao2003ion}. 

From a theoretical standpoint, the surfaces are often modeled as planar interfaces considering the large size difference between the ions and the confining macromolecules. 
In many systems under dilute electrolyte conditions, the electric field associated with the charged interfaces decays exponentially with distance in good agreement with the behavior predicted by the Debye-Huckel theory \cite{israelachvili2015intermolecular} where the characteristic decay length scales inversely with the square root of the electrolyte concentration. 
In conditions where correlations between ions are strong, e.g., for highly-concentrated electrolytes or multivalent electrolytes, an accurate description of the ionic structure requires departure from the simplest mean-field theories. Approaches such as density functional theory and integral equations where correlation effects are explicitly included take into account the important contributions of the fluctuations to the thermodynamic potential in such systems \cite{hansen2013theory,gillespie2015review,kjellander1985inhomogeneous,podgornik1989electrostatic,attard1993asymptotic,leote1994decay,ennis1995dressed,zwanikken1,jing2015ionic,kjellander2019intimate, ma2020classical}.
Computationally, the effects of correlations on the ionic structure in confinement are studied using coarse-grained models that treat the ions as finite-sized particles and replace the molecular structure of the solvent and material surfaces with dielectric continua \cite{barros2014efficient,marchi2001dielectric,messina1,attard,santos,netz2018,qin2016image,jing2015ionic,wang2017ion,lamperski2019off,mussotter2020heterogeneous}.
Some studies have also probed the effects of the solvent structure on the organization of ions near interfaces using explicit-solvent models \cite{boda2000capacitance,lamperski2015planar,allen2002intermittent,feng2010ion}. 
These investigations have shown that strong ionic correlations can have profound effects on the ionic structure, which can significantly alter the screening of the charged surfaces and change the character of the effective interaction between them.

In addition to the explicit methods of calculation, several phenomenological approaches including modified Poisson-Boltzmann theories and heuristic constructions have also been useful in developing an intuitive understanding of the effects of correlations and discussing the results of simulations describing the structural organization of ions \cite{kornyshev2007double,bazant2011double,yu2006effects,jadhao2013variational,solis2013generating}. In the context of electrolytes and their interactions with macroions, or charged surfaces, these include the notions of ion condensation \cite{manning1969limiting} and overcharging \cite{nguyen2001overcharging}, the identification of forces as induced by ion correlations \cite{rouzina1996macroion}, and the description of ion environments in terms of Wigner cells \cite{shklovskii1999wigner} or ionic glasses \cite{solis2000collapse}. 

Recent experiments have directly measured the behavior of the decay length associated with the electric field originating from the charged surface across a wide range of electrolyte solutions at high concentrations \cite{smith2016electrostatic,perez2017underscreening,gaddam2019electrostatic,hjalmarsson2017switchable,gebbie2017long,baimpos2014effect,smith2020forces}. Based on surface force measurements, Perkin and co-workers \cite{smith2016electrostatic, perez2017underscreening} showed that the decay of the electrostatic force between two charged surfaces in a highly-concentrated electrolyte solution is much weaker than the Debye-Huckel prediction. Similar ``underscreening” effect was observed by Gaddam and Ducker via an alternate approach involving the measurements of the surface excess of fluorescein \cite{gaddam2019electrostatic}. These experiments show evidence for the electrostatic decay or screening length that greatly exceeds the theoretical Debye length and rises with increase in concentration $c$ for high $c$ values. This observation has been made for a wide range of electrolytes from simple salt solutions (LiCl, NaCl, CsCl in water) to pure ionic liquids to ionic liquids dissolved in solvents, signaling the universal nature of the behavior of highly-concentrated electrolytes.

These experiments have inspired many recent theoretical investigations of highly-concentrated electrolyte systems. Efforts have been made to describe the properties of electrolytes and the non-monotonic behavior of the associated decay length using both analytical approaches \cite{goodwin2017underscreening,rotenberg2018underscreening,coupette2018screening,kjellander2019intimate,ma2020classical,de2020interfacial,bresme2018debye} and computer simulations \cite{coles2020correlation,de2020interfacial,dopke2019preferential,wang2020structural}. 
Most studies have focused on understanding the non-monotonic behavior of the decay length by computing pair correlation functions in bulk electrolyte systems \cite{coles2020correlation,rotenberg2018underscreening}. Some computational studies have focused on measuring the screening behavior of electrolytes in confined electrolyte systems  \cite{de2020interfacial,wang2017ion,wang2020structural,dopke2019preferential}.

In an earlier publication \cite{jing2015ionic}, we investigated the structure of electrolyte ions in the confinement formed near uncharged planar interfaces using coarse-grained molecular dynamics (MD) simulations of restricted primitive models of electrolytes where cations and anions are assumed to have the same size. The distribution of ions in confinement created by unpolarizable or polarizable interfaces was extracted for a fixed interfacial separation of 3 nm for different values of electrolyte concentration ($< 1$ M), ion valency, and dielectric mismatch at the interface. Ions were found to deplete from the interfaces, adsorb to the interfaces, and show density oscillations. We showed that the forces that govern these structural features are not always directly exerted by the interfaces, which were charge neutral in some cases, but arise from the thermal motion of the ions and the geometric constraints that the interfaces impose. 

In this paper, we use MD simulations to perform a systematic study of the ionic structure of aqueous monovalent electrolyte solutions confined by two planar interfaces over a wide range of concentrations $c \in (0.1, 2.5)$ M, interfacial separations $h \in (5,8)$ nm, surface charge densities $\sigma_s \in (-0.005,-0.02) \mathrm{C}/\mathrm{m}^2$, and counterion size $d_+ \in (0.2 - 0.63)$ nm. 
Our focus is on understanding the behavior of ions in confinement created by charged interfaces under high electrolyte concentrations \cite{smith2016electrostatic,gaddam2019electrostatic}. To keep the computational costs tractable for investigations over a wide range of interfacial separations and solution conditions, the primitive model for the electrolyte system is adopted. The model system of cations and anions with diameters corresponding to hydrated diameters of Na and Cl ions is comprehensively studied. To complement the MD simulation results, we also develop a variational approach based on a free energy functional that captures the key features of the electrolyte system.

In order to probe the non-monotonic behavior of the decay length observed for a diverse set of electrolytes and material surfaces, \cite{smith2016electrostatic,gaddam2019electrostatic,perez2017underscreening,gebbie2017long} the effects arising due to specific material surfaces, including the effects of surface polarization charges, are neglected in favor of examining the general interplay of electrostatics-driven ion accumulation (depletion) near charged interfaces and the effects of steric correlations arising due to finite ion size. 
Our previous paper showed that the thermal forces arising due to the steric correlations can often overwhelm the effects due to the surface polarization charges for monovalent electrolytes at high concentration ($\gtrsim 0.1$ M) \cite{jing2015ionic}.
The ionic structure is quantified by evaluating the ion number densities, net charge densities, screening factors, \cite{qiao2004charge,martin2009additional,de2020interfacial,dopke2019preferential} and the characteristic decay length associated with the screening of the charged interface. 
Simulation results show the presence of two distinct regimes of screening behavior as the concentration is increased across a wide range of systems generated by tuning the interfacial separation, surface charge density, and ion size. 
The variational approach produces ion density profiles and screening factors with characteristics consistent with those observed in simulations. The results demonstrate the relation between the appearance of strong correlations and changes in the screening behavior.  

\section{Models and Methods}
Our model system consists of an electrolyte solution confined within two charged, planar interfaces parallel to each other.
Each interface is characterized with a uniform surface charge density $\sigma_s < 0$.
The simulation cell is a rectangular box of dimensions $l_{x}\times l_{y}\times h$, with $l_x = l_y$. Periodic boundary conditions are employed in the \(x\mbox{-}\) and \(y\mbox{-}\) directions. $l_{x}$ and $l_{y}$ are chosen to be sufficiently large to avoid any artifacts due to the periodic boundary conditions.
$z = -h/2$ and $z = h/2$ planes are chosen as the location of the interfaces and $h$ is defined as the interfacial separation. The coordinate system is defined such that $z=0$ corresponds to the midpoint between the interfaces. 
The distribution of ions is examined for $5 \le h \le 8$ nm. 
The uniform charge density on the planar interfaces is simulated by meshing each interface with a large number of points $M$ and assigning each point the same charge $q = \sigma_s l_x^2 / M$.
For all systems, $M=2500$ is used. We verified that similar results for the ionic distributions were obtained with larger values of $M$.

Ions are modeled using the primitive model of electrolytes \cite{messina1,boda,jing2015ionic} with hydrated diameters. For example, a model electrolyte having cations and anions of sizes informed by the hydrated size of Na$^+$ (0.474 nm) and Cl$^-$ (0.627 nm) \cite{marcus1988ionic} is extensively studied. Water confined within the planar interfaces is modeled as an implicit solvent with dielectric permittivity of $80$.  

The Hamiltonian of the confined electrolyte system is the sum of the total steric potential energy between ions, the total electrostatic energy between ions, and the total energy associated with the interaction between the ions and the two planar interfaces. 
Ion-ion steric interactions are modeled using the standard purely-repulsive and shifted Lennard-Jones (LJ) potential \cite{barros2014efficient,jing2015ionic}. For a pair of ions separated by distance $r$, this potential is given as
\begin{equation}
\frac{U_{LJ}}{k_BT} = 1 + 4 \left[\left(\frac {d}{r}\right)^{12} - \left(\frac {d}{r}\right)^{6}\right],
\end{equation}
for $r \le 2^{1/6}d$, where, $d$ is the average diameter of the two ions, $k_B$ is the Boltzmann constant, and $T$ is the temperature. For $r > 2^{1/6}d$, $U_{LJ} = 0$. Steric interactions between an ion and the interface are modeled using a similar repulsive LJ potential with a different size parameter $d$ that represents the distance of the closest approach of an ion to the interface.
Electrostatic interactions between ions and between an ion and the interface are modeled using the Coulomb potential. The long-range of the Coulomb potential is properly treated using Ewald sums \cite{deserno1998mesh} or the method of charged sheets \cite{boda1998monte,guerrero2013enhancing,jing2015ionic}. Either approach gives similar results within statistical uncertainties. 

Simulations are performed using LAMMPS \cite{lammps.plimpton} as well as codes developed in our lab \footnote{Code available at https://github.com/softmaterialslab/nanoconfinement-md}. 
For all systems, the dimensions of the simulation box in the unconfined $x$ and $y$ directions are taken to be $l_x = l_y = 18$ nm.  
Depending on the electrolyte concentration, interfacial separation and surface charge density, the number of particles in the main simulation cell varied between 96 and 7884. 
Note that in addition to electrolyte ions, counterions are  included in the confinement to ensure electroneutrality. Counterions are modeled as positively-charged ions of the same diameter and charge as electrolyte cations.
All simulations are performed in an NVT ensemble at a temperature $T=298$ K that is maintained using a Nose–Hoover thermostat \cite{nose1984unified}. Each system is simulated for 1 ns to reach equilibrium with a timestep of $1$ femtosecond. After equilibration, systems are simulated for another 9 ns and trajectory data for computing ionic distributions is collected every 0.1 ps.

Table \ref{tab:params} shows the key model parameters and the range over which they are varied in order to probe the ionic structure associated with confined monovalent electrolytes. 
Electrolyte concentration is defined as $c = N / V$, where $N$ is the number of electrolyte cations or anions and $V = l_x \times l_y \times h$ is the volume of the simulation box. In most investigations, the anions (co-ions) are modeled as particles of diameter 0.627 nm (associated with the hydrated size of a Cl$^{-}$ ion). 

\begin{table}[t]
\begin{flushleft}
\small
\begin{tabular*}{0.48\textwidth}{@{\extracolsep{\fill}}llll}
    \hline
    Parameter & Range \\
    \hline
    Interfacial separation ($h$)   &  5 -- 8 nm \\
    Salt concentration ($c$)   & 0.1 -- 2.5 M \\
    Cation diameter ($d_+$)     & 0.209 -- 0.627 nm \\
    Surface charge density ($\sigma_{s}$) & $-$0.005 -- $-$0.02 $\mathrm{C}/\mathrm{m}^2$ \\
    \hline
  \end{tabular*}
\caption{Key parameters in the primitive model.}
\label{tab:params}
\end{flushleft}
\end{table}

\subsection{Ionic structure measurements}\label{sec:measure}
The number density in the direction perpendicular to the interfaces for cations and anions are extracted using the trajectory data collected during the simulation. The net charge density $\rho(z)$ at $z$ is calculated as:
\begin{equation}\label{eq:rhonet}
\rho(z) = en_{+}(z) - en_{-}(z),
\end{equation}
where $e$ is the electronic charge, and $n_{+}(z)$ and $n_{-}(z)$ are the number density profiles of positively-charged and negatively-charged ions respectively.
To characterize the effects of ion depletion or accumulation near the interface on the screening of the surface charge, the integrated charge or screening factor \cite{qiao2004charge} $S$ is extracted using the following equation \cite{wang2017ion,dopke2019preferential}: 
\begin{equation}\label{eq:S}
S(z) = \sigma_{s} + \int_{-h/2}^{z} \rho(z') dz'.
\end{equation}
Here, $\sigma_{s}$ is the surface charge density and $\rho$ is the charge density given by Equation \ref{eq:rhonet}. Note that $S(z)$ is defined using the left planar interface as the reference surface. For clarity, we plot $S$ vs. $z + h /2$ where the latter denotes the distance from the left interface. $S(z)$ is well defined for $-h/2 \le z \le 0$, i.e., up to the center of the confined region. 
The first term $\sigma_s$ in Equation \ref{eq:S} is added to shift the integrated charge in order to satisfy an appropriate boundary condition at the interface: $S(-h/2) = \sigma_s$. This condition implies that as the distance from the interface approaches 0, the surface is not screened and $S$ is simply the bare surface charge density $\sigma_s$.  

\begin{figure*}[tb]
\centering
\includegraphics[width=0.9\textwidth]{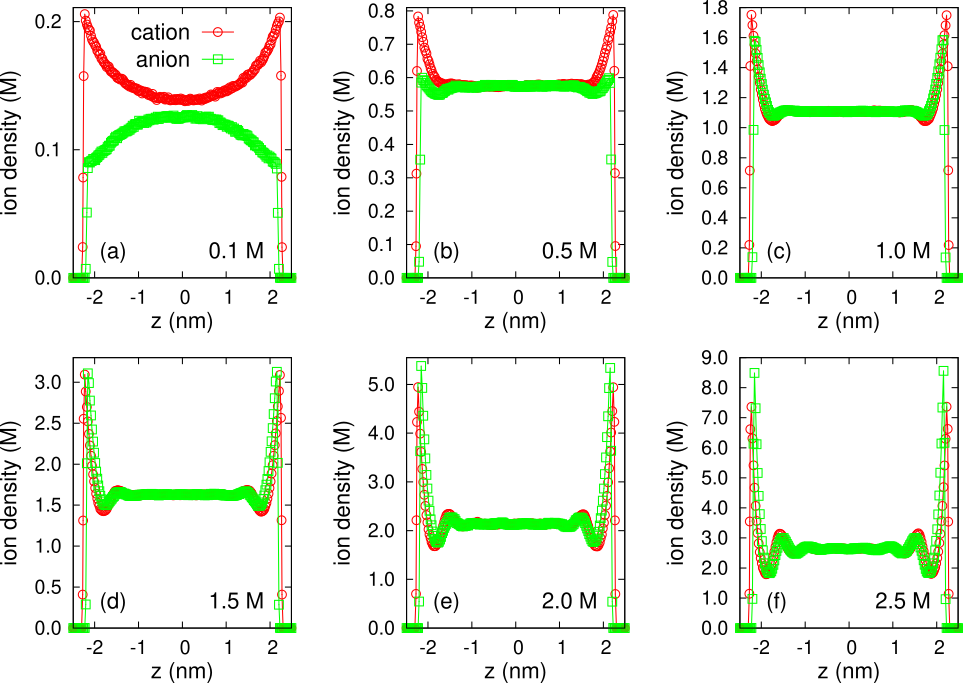}
\caption{Density profile $n_+(z)$ of cations (circles) and $n_-(z)$ of anions (squares) confined within two charged planar interfaces, each characterized with surface charge density $\sigma_{s} = -0.01 \mathrm{C}/\mathrm{m}^2$. Results are shown for different electrolyte concentrations: a) 0.1 M, b) 0.5 M, c) 1.0 M, d) 1.5 M, e) 2.0 M, f) 2.5 M. In each case, the interfaces are separated by 5 nm.}
\label{fig:rho_nacl}
\end{figure*}

$S(z)$ provides a way to measure the screening of the surface charge on the interface by the presence of electrolyte ions at a given concentration. The distance $z^*$ where $S$ decays to 0 can be interpreted as a length-scale associated with the decay of the electric field originating from the charged surface. In practice, $z^*$ is extracted as the minimum distance beyond which $|S|$ has decayed to less than $5\%$ of its value at the left interface ($\sigma_s$). 
It is useful to note that under conditions where the screening behavior is well described by Debye-Huckel theory, $S$ exhibits a simple exponential decay and the distance at which $S$ decays to less than $5\%$ of its initial value is $\approx 3\lambda_D$, where $\lambda_D$ is the Debye length.

The concept of integrated charge has been used in many studies to quantify the effects of the ionic structure on the charged surface \cite{qiao2004charge,martin2009additional,wang2017ion, dopke2019preferential,de2020interfacial, vsantic2017monte, bresme2018debye}. Instead of Equation \ref{eq:S}, $S$ is often defined by normalizing the integral (second term on the right hand side of the equation) by $\sigma_s$ such that $S$ decays to 1 instead of 0 \cite{qiao2004charge,de2020interfacial}.
We also note that in a recent study using classical density functional theory, the decay length was extracted by examining the behavior of ions near a charged surface and calculating the slope of ${\log|\rho|}$ \cite{ma2020classical}. 

\section{Results}

\subsection{Ionic density profiles} \label{sec:rho}

We first present the results for the ionic distributions extracted using simulations of the model electrolyte system comprising of cations and anions of diameters 0.474 nm and 0.627 nm respectively. These ion sizes are informed by the hydrated diameters of Na and Cl ions in water \cite{marcus1988ionic}.
Figure \ref{fig:rho_nacl} presents the density profiles $n_+(z)$ and $n_-(z)$ of cations and anions confined between planar interfaces separated by 5 nm and characterized with surface charge density $\sigma_{s} = -0.01 \mathrm{C}/\mathrm{m}^2$. Profiles are shown for different electrolyte concentrations in the range $c \in 0.1 - 2.5$ M. For each system, $n_+(z)$ and $n_-(z)$ are symmetric around $z=0$, as expected. For $c = 0.1$ M, cations (counterions) are accumulated near the planar interfaces while anions (co-ions) are depleted near the interfaces. This behavior can be attributed to the electrostatic force exerted by the negatively-charged surface. 
As $c \gtrsim 0.5$ M, both cations and anions are accumulated near the interfaces as evidenced by the peaks of $n_+(z)$ and $n_-(z)$ near the interfaces. The accumulation of anions near the interfaces can be attributed to the thermal forces arising from the stronger steric correlations between anions that overcome the repulsive electrostatic force between the anions and the charged surface.
The location of the peak of $n_-(z)$ is farther from the interface compared to that of $n_+(z)$ because of the larger size of the anions.

\begin{figure}[ht]
\centering
\includegraphics[width=0.48\textwidth]{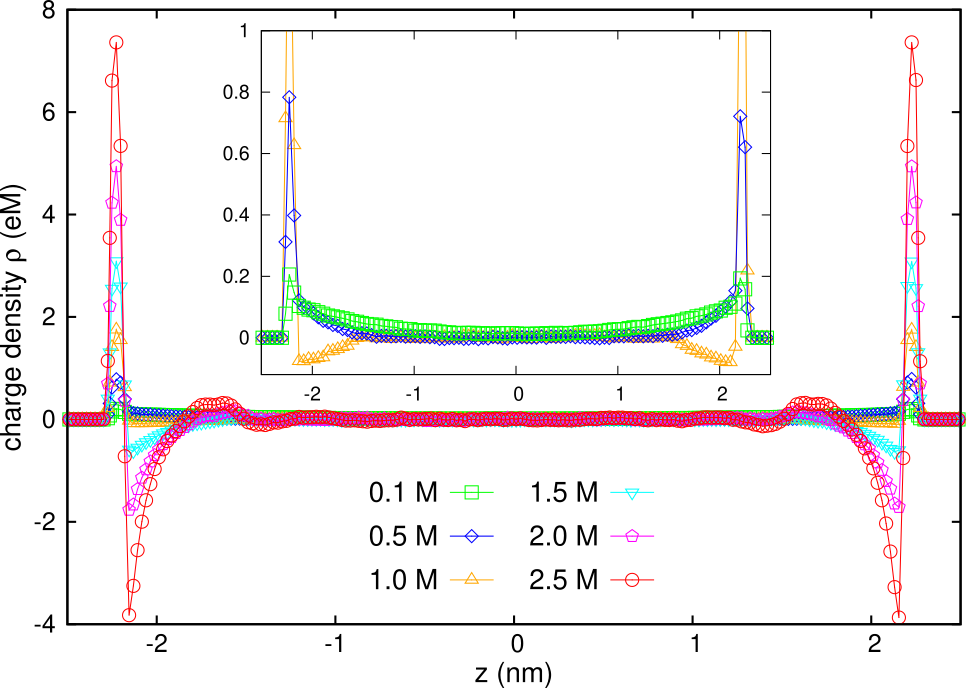}
\caption{Charge density $\rho$ for the same electrolyte systems shown in Figure \ref{fig:rho_nacl}. For clarity, the results for 0.1 M, 0.5 M, and 1.0 M for $\rho < 1e\mathrm{M}$ are shown in the inset.}
\label{fig:netrho_nacl}
\end{figure}

The accumulation of cations and anions near the interfaces continues to rise with increasing $c$. 
Figure \ref{fig:rho_nacl} (e) and (f) show that the peak of $n_-$ near the interface becomes higher than that of $n_+$ for $c \gtrsim 2$ M. This effect can be attributed to the larger size of anions compared to cations, and would disappear in size-symmetric electrolytes (see Figure \ref{fig:rho627} in the Appendix).
The near-interface accumulation of ions of both species leads to the formation of structured layers that act as soft walls and induce secondary accumulation, thus nucleating more layers. 
The cascade of soft walls resulting from the external planar interface creates oscillations in the ionic structure for $c = 2$ M and $2.5$ M.
The changes in the distribution of confined cations and anions with increasing $c$ for interfacial separation of $h=5$ nm are also observed for different $h = 6, 7, 8$ nm (Figure \ref{fig:rhoat7nm} in the Appendix shows the results for $h = 7$ nm).

We next examine the evolution in the ionic structure using the net charge density $\rho(z)$ extracted from the associated number densities $n_+(z)$ and $n_-(z)$ by employing Equation \ref{eq:rhonet}. Figure \ref{fig:netrho_nacl} shows $\rho(z)$ for the same systems explored in Figure \ref{fig:rho_nacl}. 
$\rho$ is symmetric around the center of the confinement ($z=0$). 
For the ease of exposition, we discuss the behavior of $\rho$ near the left interface.
In the immediate vicinity of the interface, $\rho$ is positive and exhibits a peak close to the interface which can be attributed to the accumulation of cations (counterions). The height of this peak increases with increasing $c$ from 0.1 M to 2.5 M. For $c \lesssim 0.5$ M, $\rho$ exhibits a monotonic decay to 0 as $z$ approaches 0 (Figure \ref{fig:netrho_nacl} inset). 
For $c \sim 1$ M, a region close to the interface develops where $\rho$ becomes negative reaching a minimum of $\approx -0.1e$M before decaying to 0. This non-monotonic behavior can be attributed to the gradual increase in the accumulation of anions (co-ions) near the interface with increasing concentration. 

\begin{figure}[ht]
\centering
\includegraphics[width=0.48\textwidth]{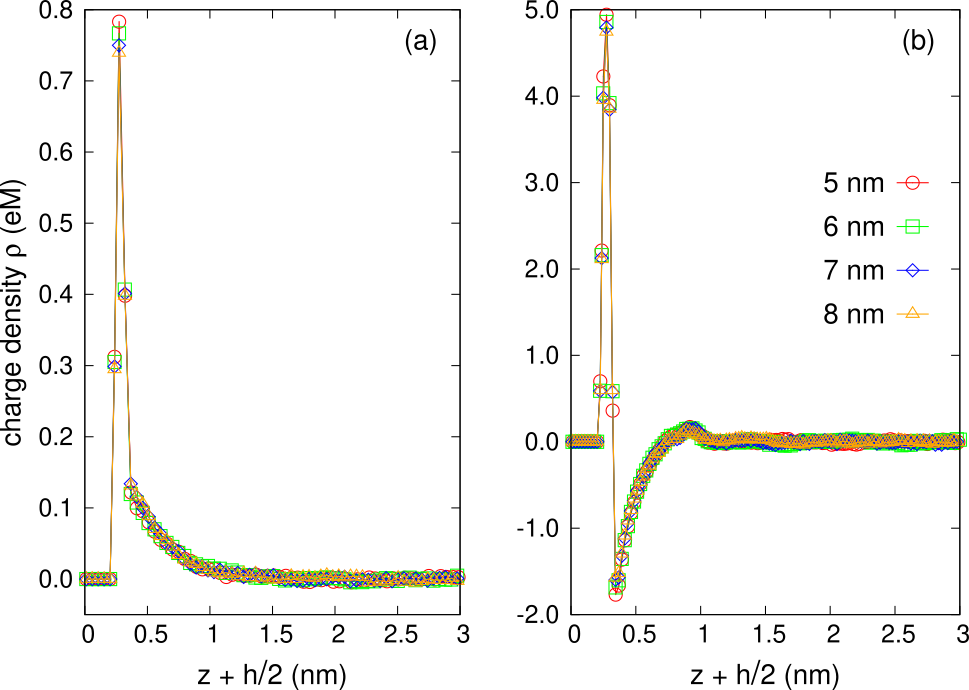}
\caption{Charge density $\rho$ as a function of the distance $z + h/2$ from the left interface for different interfacial separations $h = $ 5 nm (circles), 6 nm (squares), 7 nm (diamonds), and 8 nm (triangles). Results are shown for electrolyte concentration $c$ = 0.5 M (a) and 2.0 M (b). $\rho$ exhibits a similar behavior for different $h$. For low $c$, after the initial peak, $\rho$ shows a monotonic decay to 0 as $z+h/2$ is increased. In contrast, for high $c$, $\rho$ exhibits a non-monotonic behavior before decaying to 0.}
\label{fig:rho_diff_conf}
\end{figure} 

As $c$ rises to values $\gtrsim 2$ M, a prominent $\rho < 0$ region develops adjacent to the first positive peak close to the interface and the minimum value of $\rho$ drops sharply.
The drop can be attributed to the strong accumulation of anions near the interface seen in Figure \ref{fig:rho_nacl} (e) and (f). The excess net negative charge associated with the layer of anions induces a second positive peak adjacent to it, which in turn nucleates a second $\rho < 0$ region farther from the interface (seen more clearly for $c=2.5$ M in Figure \ref{fig:netrho_nacl}). The resulting oscillatory, non-monotonic behavior of $\rho$ is associated with the structural modulations present in $n_+$ and $n_-$ seen in Figure \ref{fig:rho_nacl}.
The amplitude of the oscillations is reduced and $\rho$ gradually decays to 0 as $z$ approaches the center of the confinement ($z \to 0$).

It is useful to analyze the charge density $\rho$ for different interfacial separations at a given electrolyte concentration. 
Figure \ref{fig:rho_diff_conf} shows $\rho$ as a function of the distance $z + h/2$ from the interface for $h = 5,6,7,8$ nm for $c = 0.5$ M (a) and $2.0$ M (b). By symmetry considerations, the distance is measured relative to the left interface. 
For both low and high electrolyte concentrations, the data for $\rho$ under different $h = 5, 6, 7, 8$ nm falls on the same universal curve. For $c = 0.5$ M, $\rho$ exhibits an initial peak near the interface and then decays monotonically to 0 as $z$ is increased. For higher $c = 2$ M, $\rho$ exhibits a non-monotonic, oscillatory behavior (with a prominent $\rho < 0$ region) before decaying to 0 for larger $z+h/2$. The distinct behavior of $\rho$ under low and high $c$ conditions is similar to the trend observed in Figure \ref{fig:netrho_nacl}.

\subsection{Screening factor}\label{sec:sf}

\begin{figure}[t] 
\centering
\includegraphics[width=0.48\textwidth]{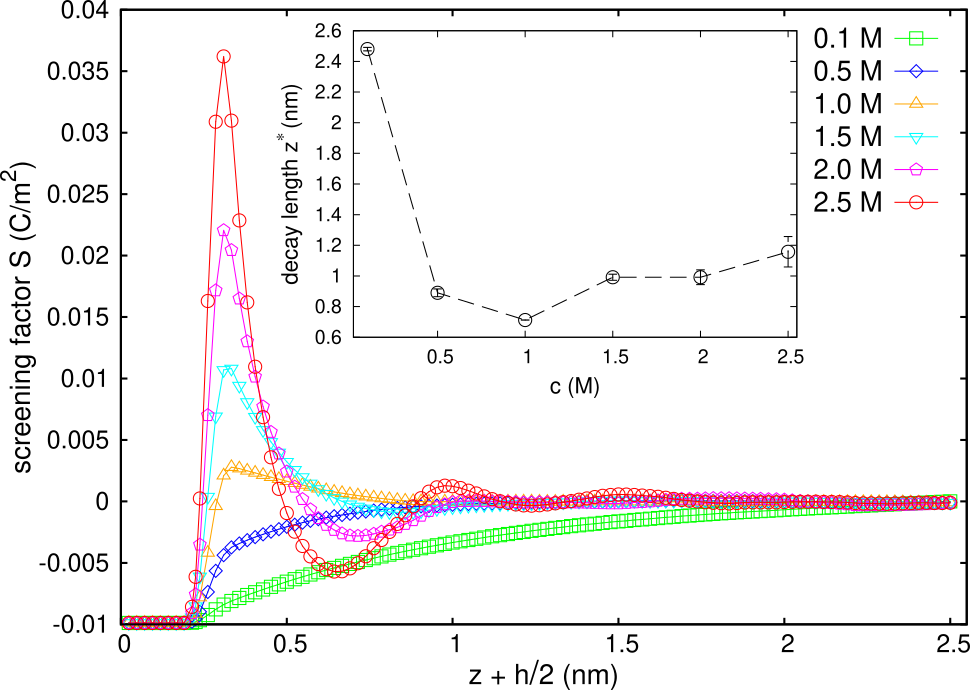}
\caption{Screening factor $S$ vs. the distance $z+h/2$ from the left interface for the electrolyte systems shown in Figure \ref{fig:rho_nacl}. Inset shows the decay length $z^*$ extracted as the distance where $S$ decays to $\lesssim 5 \%$ of its value at $z + h/2 = 0$. $z^*$ is the average of the decay lengths obtained for different $h = 5, 6, 7, 8$ nm.}
\label{fig:sf_0474lowchargesurface}
\end{figure}

We next analyze the screening factor $S(z)$ as a function of the distance $z + h/2$ from the left planar interface for different electrolyte concentrations. $S$ is extracted using Equation \ref{eq:S} by employing the results for the net charge density $\rho(z)$ shown in Figure \ref{fig:netrho_nacl} for an interfacial separation of $h=5$ nm. 
Figure \ref{fig:sf_0474lowchargesurface} shows $S(z)$ vs. $z+h/2$ for the same systems studied in Figures \ref{fig:rho_nacl} and \ref{fig:netrho_nacl}. 
For all concentrations $c \in (0.1, 2.5)$ M, $S = \sigma_s = -0.01 \mathrm{C}/\mathrm{m}^2$ when $z + h/2 \lesssim d_{+} / 2$. In other words, the surface charge on the interface is unscreened up to a distance of the closest approach of the cation. 
For all $c$, $S(z)$ exhibits an initial rise as the distance $z+h/2$ is increased, implying the reduction in the integrated charge due to the screening of the surface by the electrolyte ions.
For $c \lesssim 0.5$ M, $S(z)$ exhibits a monotonic rise before decaying to 0 as the distance $z+h/2$ from the interface becomes large. Under these conditions, $S \le 0$ for all $z$, showing no evidence for any charge inversion.
For $c = 1$ M, $S$ increases sharply with $z$, reaching a value of 0 at a distance $z + h/2 \approx 0.3$ nm away from the interface. $S$ continues to rise further peaking at a value of $\approx 0.003 \mathrm{C}/\mathrm{m}^2$ ($\approx 0.3|\sigma_s|$) before decaying to 0. The presence of $S(z) \gtrsim 0$ region for $z + h/2 \gtrsim 0.3$ nm suggests a weak charge inversion.

For $c \gtrsim 1.5$ M, $S$ continues to rise rapidly to values beyond 0 reaching much higher peaks (e.g., $>0.03 \mathrm{C}/\mathrm{m}^2$ for 2.5 M), signaling greater charge inversion. More importantly, instead of decaying smoothly down to 0 (like for $c = 1$ M), $S$ exhibits a non-monotonic, oscillatory behavior and becomes negative with increasing $z + h/2$, exhibiting a minimum before decaying to 0. The oscillations are stronger for $c=2.5$ M, where a clear secondary peak $\approx 1$ nm away from the interface is observed.
Similar results for $S(z)$ are obtained for electrolytes confined under different interfacial separations $h = 6, 7, 8$ nm. This finding is expected given the variation of $\rho$ as a function of $z + h/2$ for different $h$ (Figure \ref{fig:rho_diff_conf}). Figure \ref{fig:sf_7nm} in the Appendix shows $S$ for $h=7$ nm.

\begin{figure}[th]
\centering
\includegraphics[width=0.48\textwidth]{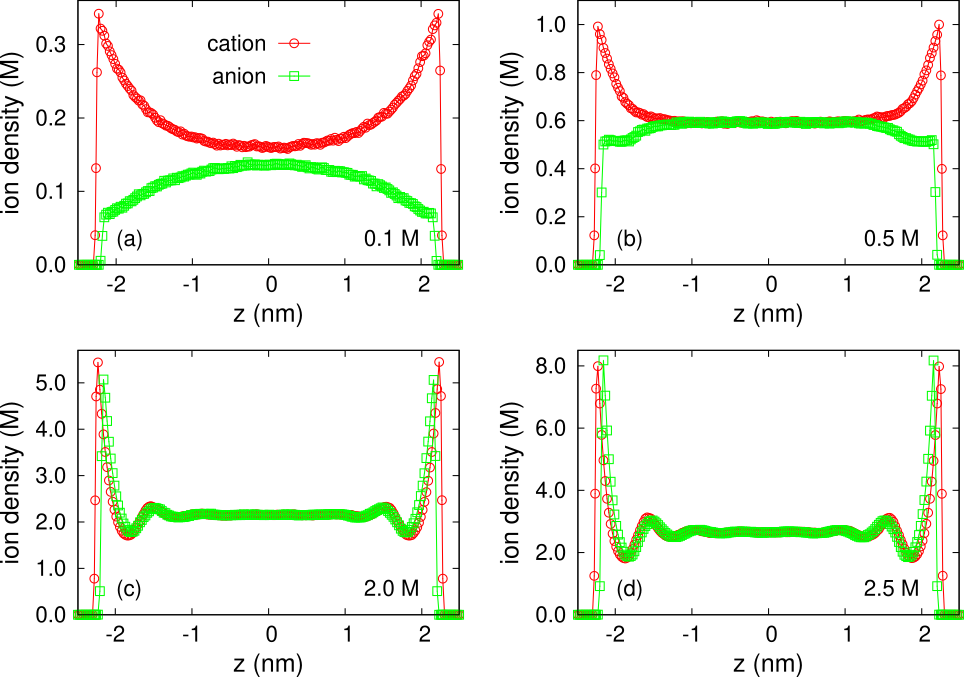}
\caption{Distributions of cations (circles) of diameter 0.474 nm and anions (squares) of diameter 0.627 nm confined by interfaces characterized with surface charge density $-0.02 \mathrm{C}/\mathrm{m}^2$ and separated by 5 nm. Results are shown for different electrolyte concentrations: a) 0.1 M, b) 0.5 M, c) 2 M, d) 2.5 M.}
\label{fig:rho_highsigma}
\end{figure}

Using $S(z)$, we extract the decay length $z^*$ following the procedure described in Section \ref{sec:measure}. 
Recall that $z^*$ is the minimum distance beyond which $|S|$ is less than approximately $5\%$ of its value at the left interface. $z^*$ acts as a measure of the extent of the screening of the charged surface by the electrolyte ions. 
The availability of data for $S(z)$ vs. $z + h/2$ for many $h$ values enables an accurate assessment for $z^*$, providing also the associated statistical uncertainties. $z^*$ is computed by taking the average of the decay lengths obtained for different $h = 5, 6, 7, 8$ nm.
The inset in Figure \ref{fig:sf_0474lowchargesurface} shows $z^*$ vs. $c$.
The error bars are computed by extracting the standard deviation associated with the $z^*$ data obtained for different $h = 5, 6, 7, 8$ nm.
We find that $z^*$ decreases sharply as $c$ is increased up to $\approx 1$ M. However, for $c \gtrsim 1$ M, $z^*$ increases slightly with increasing $c$.

\subsection{Effects of changing surface charge density}

The results shown thus far have been obtained for planar interfaces characterized with surface charge density $\sigma_s = -0.01\mathrm{C}/\mathrm{m}^2$. 
It is useful to assess how the distinct ionic structure observed under low and high electrolyte concentrations is affected by changing $\sigma_s$.
We performed simulations of the same model electrolyte system confined by interfaces for 2 different values of $\sigma_s = -0.005, -0.02 \mathrm{C}/\mathrm{m}^2$. Changing $\sigma_s$ alters the distributions of ions in confinement, however, the distinct features associated with the ionic structure separating the two regimes of low $c$ ($\lesssim 1$ M) and high $c$ ($> 1$ M) persist for different $\sigma_s$ values. Figure \ref{fig:rho_highsigma} shows the distribution of cations and anions confined by surfaces characterized with $\sigma_{s} = -0.02 \mathrm{C}/\mathrm{m}^2$ for $c \in (0.1, 2.5)$ M. We find that the larger negative charge on the interfaces leads to enhanced accumulation of positive ions and stronger depletion of anions compared to the $\sigma_s = -0.01 \mathrm{C}/\mathrm{m}^2$ case. For example, unlike the behavior observed in Figure \ref{fig:rho_nacl}(b), anions do not exhibit a clear accumulation near the interfaces for $c=0.5$ M. Further, even at the highest $c = 2.5$ M, the peak associated with the anions does not significantly exceed the cation peak, as was the case for $\sigma_s = -0.01 \mathrm{C}/\mathrm{m}^2$ for $c\gtrsim 2$ M (Figure \ref{fig:rho_nacl}(e) and (f)).

\begin{figure}[t]
\centering
\includegraphics[width=0.48\textwidth]{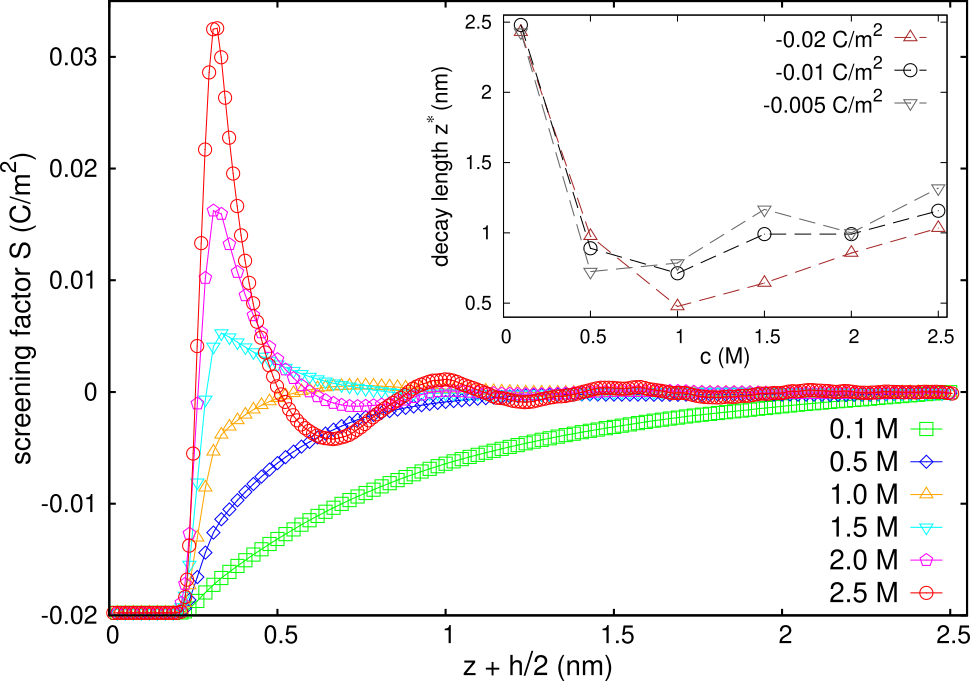}
\caption{Screening factor $S(z)$ vs. the distance $z+h/2$ from the left interface for electrolytes with cations of diameter 0.474 nm and anions of diameter 0.627 nm confined by planar interfaces separated by 5 nm and characterized with surface charge density $\sigma_{s} = -0.02 \mathrm{C}/\mathrm{m}^2$. Results are shown for different electrolyte concentrations $c \in (0.1,2.5)$ M. 
Inset shows the decay length $z^*$ vs. $c$ for interfaces characterized with different $\sigma_s = -0.005\mathrm{C}/\mathrm{m}^2, -0.01\mathrm{C}/\mathrm{m}^2, -0.02 \mathrm{C}/\mathrm{m}^2$.
}
\label{fig:sf_0474highchargesurface}
\end{figure}

Figure \ref{fig:sf_0474highchargesurface} shows the screening factor $S(z)$ as a function of the distance $z+h/2$ from the left interface characterized with $\sigma_{s}$ = $-0.02 \mathrm{C}/\mathrm{m}^2$ for different $c \in (0.1, 2.5)$ M. 
Quantitative differences are observed in $S(z)$ compared to the results for the system with $\sigma_{s}$ = $-0.01 \mathrm{C}/\mathrm{m}^2$ (Figure \ref{fig:sf_0474lowchargesurface}). However, 
the overall $S$ behavior at low $c$ is qualitatively distinct from the behavior at high $c$, consistent with the smaller $\sigma_s$ results.
When the concentration is low ($c \lesssim 1$ M), $S(z) \le 0$ and exhibits a monotonous rise before decaying to 0 for large $z+h/2$. 
The screening behavior for high $c \gtrsim 1.5$ M is entirely different. Similar to $S(z)$ for $\sigma_s = -0.01 \mathrm{C}/\mathrm{m}^2$ (Figure \ref{fig:sf_0474lowchargesurface}), $S$ rises rapidly to values $> 0$ and then exhibits a non-monotonic, oscillatory decay to 0 for large $z+h/2$. Similar results are observed for the case of $\sigma_s = -0.005 \mathrm{C}/\mathrm{m}^2$.
Figure \ref{fig:sf_0474highchargesurface} (inset) shows $z^*$ as a function of electrolyte concentration $c$ for different $\sigma_s$ values. We find that $z^*$ vs. $c$ for $\sigma_{s}$ = $-0.005, -0.02 \mathrm{C}/\mathrm{m}^2$ exhibits a similar behavior as the result for $\sigma_{s}$ = $-0.01 \mathrm{C}/\mathrm{m}^2$. 
$z^*$ decreases as $c$ rises up to $\sim 1$ M. For $c > 1.0$ M, $z^*$ rises with increasing $c$. 

\begin{figure}[ht]
\centering
\includegraphics[width=0.48\textwidth]{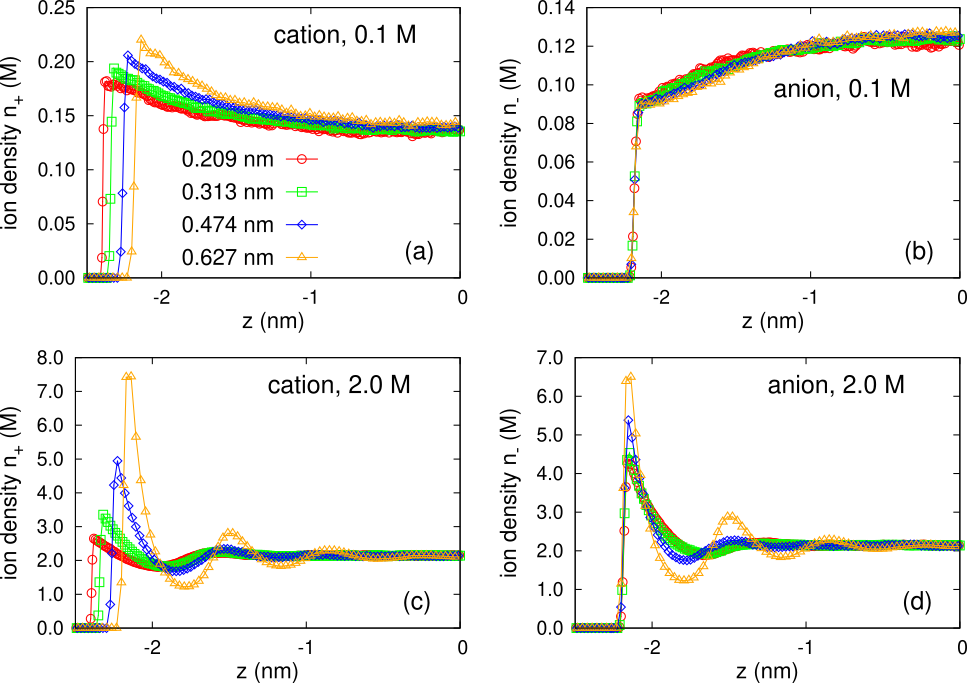}
\caption{Distributions of cations and anions confined within interfaces separated by 5 nm and characterized with a surface charge density $\sigma_{s}$ = $-0.01 \mathrm{C}/\mathrm{m}^2$. Different symbols correspond to electrolytes with cations of different diameters $d_+ = 0.209$ nm, $0.313$ nm, $0.474$ nm, $0.627$ nm. All electrolytes have anions of size $d_- = 0.627$ nm. Top row shows the density profiles of cations (a) and anions (b) at 0.1 M. Bottom row shows the density profiles of cations (c) and anions (d) at 2.0 M. Results are shown for only the left region of the confinement. 
}
\label{fig:rho_size}
\end{figure}

\subsection{Effects of changing ion size}

We now investigate the effects of changing the ion size on the distributions of ions confined by planar interfaces for different electrolyte concentrations. Our focus is on assessing how the changes in the steric correlations influence the observed distinct regimes of screening behavior under low and high $c$ conditions. The diameter of the anion is fixed at $d_- = 0.627$ nm. We perform simulations of different electrolyte systems generated by changing cation diameter to $d_+ = 0.209$ nm, $0.313$ nm, and $0.627$ nm. 
Other parameters are the same as in the model electrolyte system studied in Section \ref{sec:rho}. Simulations are performed for interfaces characterized with a surface charge density $\sigma_{s}$ = $-0.01 \mathrm{C}/\mathrm{m}^2$ and for different interfacial separations $h = 5,6,7,8$ nm. We also simulate a  size-symmetric electrolyte with smaller-sized ions of diameter $d_+ = d_- = 0.313$ nm as a reference system to assess the contributions of stronger steric effects.

Figure \ref{fig:rho_size} shows the density profiles of cations and anions near the left interface for systems with cations of diameter $d_+ = 0.209$ nm, $0.313$ nm, $0.474$ nm, and $0.627$ nm under low $c = 0.1$ M (top row) and high $c = 2$ M conditions (bottom row). 
\begin{figure}[th]
\centering
\includegraphics[width=0.48\textwidth]{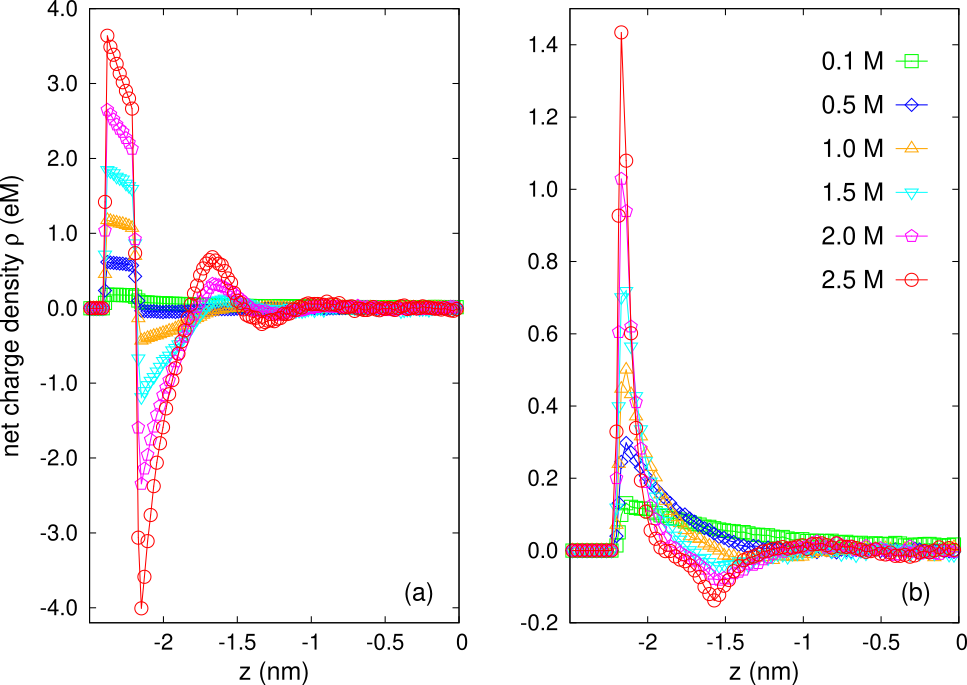}
\caption{Net charge density $\rho$ for confined electrolytes with cations of size 0.209 nm (a) and 0.627 nm (b) under different electrolyte concentrations $c \in (0.1,2.5)$ M. Both systems have anions with the same diameter (0.627 nm). Electrolytes are confined by planar interfaces characterized with a surface charge density $\sigma_{s}$ = $-0.01 \mathrm{C}/\mathrm{m}^2$ and separated by 5 nm.}
\label{fig:netrho_ionsize}
\end{figure}
For $c = 0.1$ M, as $d_+$ increases, cations accumulate farther from the interface due to the increase in the excluded volume (Figure \ref{fig:rho_size} (a)). Also, their peak density increases slightly with increasing $d_+$. The distribution of anions is weakly affected by changes in the cation size (Figure \ref{fig:rho_size} (b)). In all cases, anions deplete from the interfaces, and the depletion is slightly stronger for larger $d_+$. This can be attributed to a more effective screening of the surface charge by smaller-sized counterions as they can accumulate closer to the interface, which decreases the electrostatic force with which the anions are repelled by the interface.

Changing cation size has more dramatic effects for $c = 2$ M. Here, again due to the increase in the excluded volume, cations accumulate farther from the interface as $d_+$ increases (Figure \ref{fig:rho_size} (c)). However, the peak density of cations rises much more rapidly with cation size (from $\approx 2.8$ M at $d_+ = 0.209$ nm to $\approx 7.5$ M at $d_+= 0.627$ nm). This can be attributed to an overall increase in the ion-ion steric correlations, which push the ions towards the interface (similar to the case where the ion density near the interface rises because of increasing $c$). More importantly, the distribution of anions is strongly affected by the changes in cation size. Contrary to the trend observed for $c=0.1$ M, the peak anion density increases significantly with increasing $d_+$ at $c=2$ M (Figure \ref{fig:rho_size} (d)). Further, as $d_+$ increases, the density of anions exhibits more modulations. In Figure \ref{fig:rho_size} (d), all anions have the same size and all electrolytes are at the same concentration (2 M). The dramatic changes in the ionic structure seen here arise due to the strong steric correlations between cations and anions under high $c$ conditions. 

\begin{figure}[ht] 
\centering
\includegraphics[width=0.48\textwidth]{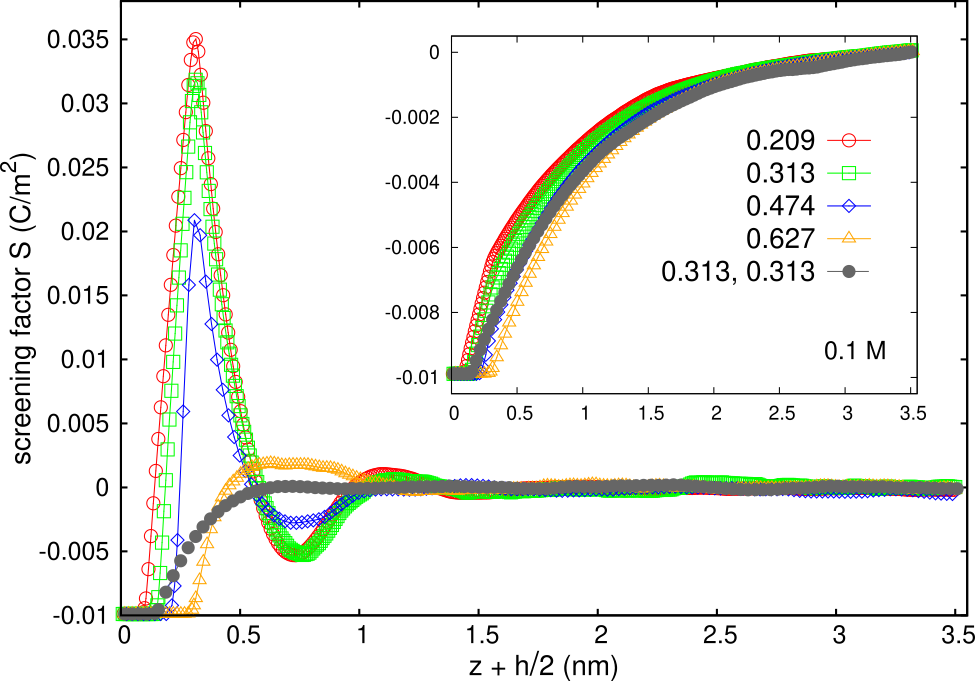}
\caption{Screening factor $S(z)$ vs. the distance $z+h/2$ from the left interface for electrolytes confined by two interfaces separated by 7 nm and characterized with surface charge density $\sigma_{s} = -0.02 \mathrm{C}/\mathrm{m}^2$. Open symbols show results for electrolytes with cations having different diameters $d_+ = 0.209$ nm, $0.313$ nm, $0.474$ nm, $0.627$ nm and anions having the same size (0.627 nm). Closed symbols are results for the size-symmetric electrolyte system having smaller-sized ions of diameter $0.313$ nm. Outset and inset show the results for electrolyte concentration $c = 2.0$ M and 0.1 M respectively.}
\label{fig:nacl_size_effect}
\end{figure}

It is instructive to examine the changes in the charge density $\rho$ vs. concentration $c$ for electrolytes with different cation sizes. Figure \ref{fig:netrho_ionsize}(a) and (b) show $\rho$ near the left interface for electrolytes with cations of diameter $d_+= 0.209$ nm and $d_+ = 0.627$ nm respectively. Results are shown for electrolytes confined within interfaces separated by $h=5$ nm. $\rho$ is derived using the cation and anion distributions associated with these systems (Figures \ref {fig:rho209} and \ref {fig:rho627} in the Appendix) for $c \in (0.1,2.5)$ M. The overall behavior of $\rho$ vs. $c$ is similar to that observed in Figure \ref{fig:netrho_nacl} for the electrolyte with cations of size $d_+=0.474$ nm. $\rho$ exhibits a transition from a monotonic behavior for sufficiently low $c$ to a non-monotonic, oscillatory behavior for high $c$. For the electrolyte system with smaller-sized cations, this transition occurs at a lower $c$ ($\sim 0.5$ M) compared to the case of electrolytes with cations of size 0.474 nm. The modulations in the ionic structure at high $c$ for the electrolyte system with cations of size $d_+ = 0.627$ nm are found to be relatively milder, which can be attributed to the size-symmetric nature of this electrolyte. Similar evolution in $\rho$ with changes in $c$ is observed for systems confined under different interfacial separations $h = 6,7,8$ nm.

We now examine how the variations in the ionic distributions resulting from changing $d_+$ affect the screening of the surface charge by the electrolyte solution. Figure \ref{fig:nacl_size_effect} shows the screening factors $S(z)$ vs. the distance $z+h/2$ from the left interface for electrolytes confined by the interfaces at $c = 2$ M and 0.1 M (inset). The interfacial separation is $h=7$ nm. Open symbols correspond to results for electrolytes with anions of fixed size $d_- = 0.627$ nm and cations of different sizes $d_+ = 0.209$ nm, $0.313$ nm, $0.474$ nm, and $0.627$ nm. Closed symbols are the results for the size-symmetric electrolyte system with ions of diameter $d_+ = d_- = 0.313$ nm.

Figure \ref{fig:nacl_size_effect} (inset) shows that when the concentration is low ($c = 0.1$ M), $S \le 0$ and exhibits a monotonous rise towards saturation to 0 as $z+h/2$ increases for all systems. 
The behavior at high $c = 2$ M is entirely different  (except for the case of the size-symmetric electrolyte with ions of diameter 0.313 nm; see below). $S(z)$ is non-monotonous and exhibits charge inversion for all $d_+$. 
For $d_+ = 0.209$ nm, $0.313$ nm, and $0.474$ nm, $S$ rises rapidly to values $> 0$, exhibits a peak, and then decays to 0 in an oscillatory fashion. 
The maximum peak of $S$ decreases as $d_+$ increases from 0.209 nm to 0.474 nm, however, its location remains the same (at $\approx 0.4$ nm).
For $d_+=0.627$ nm, $S$ rises much more slowly to values $>0$ and exhibits a smaller and broader peak, before decaying to 0 with relatively milder oscillations. This can be attributed to the size-symmetric nature of this electrolyte ($d_+ = d_- = 0.627$ nm).
Overall, these results indicate that the distinct nature of the ionic structure and the screening behavior for low and high $c$ conditions are present over a wide range of cation sizes for electrolytes with anions of size 0.627 nm. 

For the size-symmetric electrolyte with smaller-sized ions of diameter $d_+ = d_- = 0.313$ nm, $S(z)$ behavior under low and high $c$ conditions is similar. We find that even at $c=2$ M, $S$ exhibits a monotonic rise with increasing $z + h/2$ before decaying to 0 for large $z+h/2$. For this system, there is no evidence of two distinct regimes of different ionic structure and screening behavior within $c \in (0.1,2.5)$ M. We expect that the non-monotonic behavior in $S$ for this particular system appears at large $c > 2.5$ M where steric correlations become strong enough to produce structured layers of ions near the interfaces. 

\begin{figure}[ht] 
\centering
\includegraphics[width=0.48\textwidth]{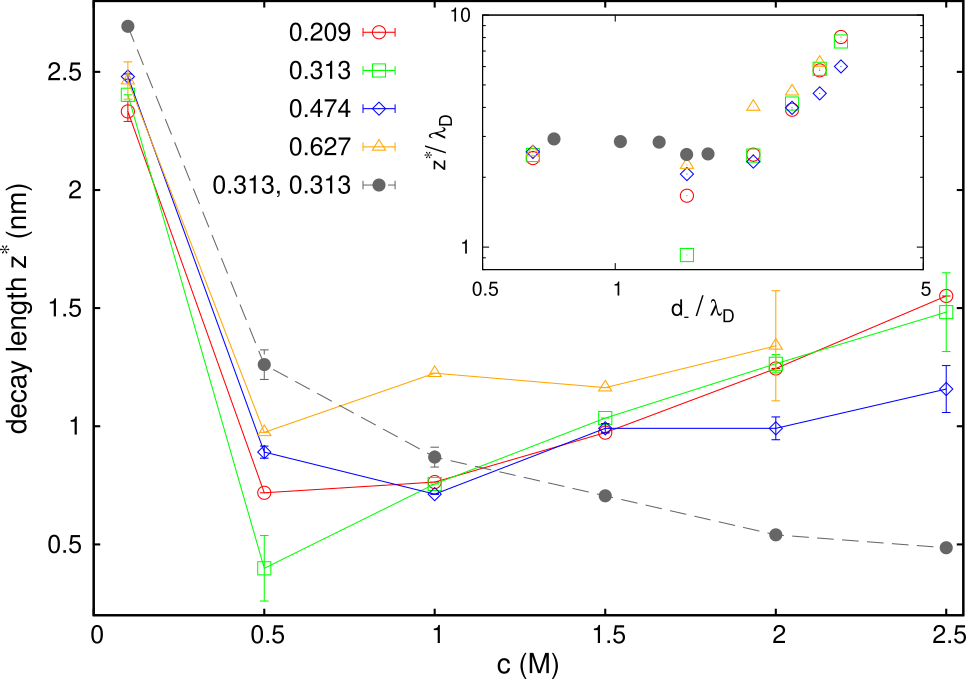}
\caption{Decay length $z^*$ vs. concentration $c$ for electrolyte systems shown in Figure \ref{fig:nacl_size_effect}. For the size-symmetric electrolyte (dashed line) with smaller-sized ions of diameter $d_+ = d_- = 0.313$ nm, $z^*$ exhibits a monotonic decrease with increasing $c$ for $c \in (0.1,2.5)$ M. For all other systems, with anions of diameter $d_-=0.627$ nm and cations having diameter $d_+ \in (0.209,0.627)$ nm, $z^*$ vs. $c$ behavior is non-monotonic. A sharp initial drop in $z^*$ with increasing $c$ is followed by a mild rise in $z^*$ as $c$ is further increased. Inset shows the log-log plot of $z^* / \lambda_D$ vs. $d_- / \lambda_D$, where $\lambda_D$ is the Debye length.}
\label{fig:decay_diff_size}
\end{figure}

Figure \ref{fig:decay_diff_size} shows the decay length $z^*$ as a function of the concentration $c$ for the different systems shown in Figure \ref{fig:nacl_size_effect}. $z^*$ is computed as an average of the results of the decay lengths extracted using $S(z)$ data for different interfacial separations $h \in (5,8)$ nm. We ensure that data from sufficiently large $h$ are employed to enable a meaningful extraction of the decay length, in particular, when the latter is large. 
$z^*$ decreases sharply for all cases as $c$ is increased up to 0.5 M. 
For electrolytes with anions of diameter $d_-=0.627$ nm and cations of diameter $d_+ \in (0.209,0.627)$ nm, $z^*$ vs. $c$ behavior is non-monotonic. 
For these systems as $c \gtrsim 1$ M, $z^*$ does not decrease with increasing $c$, but rises slightly.
The concentration where the behavior switches from sharp fall to mild rise exhibits a dependence on the ion size.
For example, the crossover $c$ for electrolyte with $d_+ = 0.627$ nm is $\approx 0.5$ M compared to $\approx 1$ M for the system with cations of size $0.474$ nm. 
In strike contrast, for the size-symmetric electrolyte with smaller ions of diameter $0.313$ nm, $z^*$ exhibits a monotonic decrease with increasing $c$ for $c \in (0.1,2.5)$ M. We hypothesize that for this system, the crossover concentration is $> 2.5$ M. 

Inspired by other studies \cite{smith2016electrostatic,coles2020correlation}, an attempt is made to collapse the data from different systems by scaling $z^*$ with $\lambda_D$ and plotting it against the scaled concentration represented as $d_- / \lambda_D$, where $\lambda_D = 1/\sqrt{8\pi l_B c}$ is the Debye length. 
The inset in Figure \ref{fig:decay_diff_size} shows the log-log plot of $z^* / \lambda_D$ vs. $d_- / \lambda_D$.
Two distinct regimes are observed. For $d_- / \lambda_D \lesssim 2$ (small $c$), the scaled decay length is roughly constant with a value of $\approx 3$. In other words, the screening of the charged surface follows the behavior predicted by the Debye-Huckel theory under these conditions. Recall that the decay length is the distance where $S$ decays to $\lesssim 5\%$ of the interfacial value, which is $\approx 3\lambda_D$ for conditions where the Debye-Huckel description is applicable. 
For $d_- / \lambda_D \gtrsim 2$ (large $c$), $z^* / \lambda_D$ rises with increasing $c$. We find this rise exhibits a power-law behavior: $z^* / \lambda_D = (d_- / \lambda_D)^n$ with exponent $n \sim 1.5$, although we realize that this scaling is observed over a comparatively limited range of data \cite{smith2016electrostatic,coles2020correlation}.

\begin{figure}[ht]
	\centering
	\includegraphics[width=0.4\textwidth]{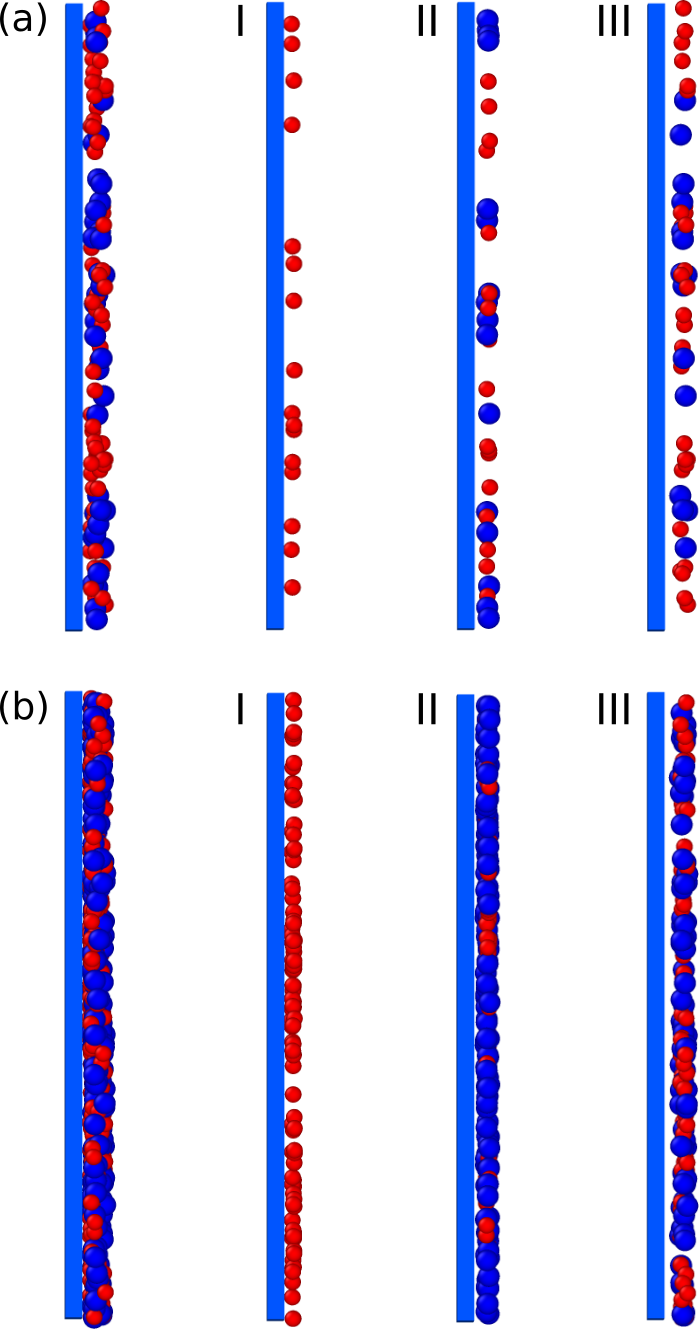}
	\caption{Representative snapshots of ions near the left planar interface extracted from the simulations of electrolytes at $c = 0.5$ M (a) and $c = 2.5$ M (b). Cations (red) and anions (blue) have diameters of 0.474 nm and 0.627 nm respectively, and the interface is characterized with surface charge density $\sigma_s = -0.02 \mathrm{C}/\mathrm{m}^2$. For both cases, the leftmost images represent the ionic structure within a distance of $\approx 0.6$ nm from the left interface. The next three images from left to right represent the features in the different layers extracted by taking thin slices of the leftmost images (see text for details).}
	\label{fig:snapshots}
\end{figure}

\subsection{Phenomenological model}

Simulation results point to the emergence of structured layers of ions near the interface and associated distinct screening behavior when the electrolyte concentration $c$ is increased. Figure \ref{fig:snapshots} (a) and (b) show representative simulation snapshots of ions near the left interface for electrolyte concentration of $0.5$ M and $2.5$ M respectively. The images shown correspond to the electrolyte system of cations (red) and anions (blue) of size 0.474 nm and 0.627 nm respectively confined by interfaces characterized with surface charge density $\sigma_s = -0.02 \mathrm{C}/\mathrm{m}^2$.
The leftmost images in (a) and (b) show ions whose centers are within a distance of $\approx 0.6$ nm from the left interface. The next three images from left to right represent the features in the different layers extracted by taking thin slices of the leftmost images. 
Slice I shows ions within $\approx 0.3$ nm distance from the interface. Slice II shows ions within a thin layer of width $\approx 0.1$ nm, bordering the slice I to its left. Slice III shows ions within a layer of width $\approx 0.2$ nm, bordering the slice II to its left. 

Slice I in either system is dominated by counterions; the number of counterions are greater for the high $c$ system (b) compared to the low $c$ system (a). Slice II associated with the high $c$ system exhibits a much larger population of co-ions (anions) compared to cations in contrast with the low $c$ case where a similar number of cations and anions are observed. 
Images associated with slice III show that farther from the interface, regions have a mixed population of cations and anions, with the high $c$ system exhibiting a more packed arrangement of ions compared to the sparse layer of ions for the low $c$ system. These snapshots are consistent with the quantitative plots shown above capturing the differences in the ionic structure under low and high $c$ conditions.

It is possible to gain insight on the simulation results using a simple mean free energy model where key features of the electrolyte system emerge naturally. A variational scheme is employed, where two distinct types of behavior are considered. The first type admits a Poisson-Boltzmann description based on a regular solution model entropy for the ions in solution. A second option considers that ions near the charged interfaces appear in highly structured distributions, forming soft aggregates (similar to the visualizations shown in Figure \ref{fig:snapshots} (b)). 

We consider a semi-infinite electrolyte with a single, charged planar interface. Near the interface, the system can acquire a spatially varying net charge density $\rho$. This charge density, along with external charges, produces a mean field potential $\phi$. The contribution to the energy of the system is then the integral of $\rho \phi/2$. Schematically, the free energy can be written as an integral over the sum of the mean field electrostatic energy and a functional density $f_\ell$ describing the local behavior of the ions:
\begin{equation}
F=\int\left[\frac{1}{2}\rho\phi+ f_\ell\right]\,dV.
\end{equation}
The functional $f_\ell$ has different forms according to the location considered.
Taking the interface to be at $z=0$, we define a near-interface region, $0<z\leq Ld$, extending $L$ ion diameters into the bulk. For simplicity, we have considered cations and anions of the same diameter $d$. In this region, the functional is $f_\ell=f_s$, where $f_s$ is constructed using a description of the ions assembled into soft aggregates. Beyond this region, where $z > Ld$, we use $f_\ell=f_b$. This second functional $f_b$ assumes a regular solution model for the entropy of the ions. 

In the variational scheme we consider, the near-interface region is not determined from the outset, and can have zero thickness, in which case the whole system is described by the bulk properties. The thickness of this layer is an outcome of the analysis of the model and depends on the bulk concentration and other parameters.  
Both explicit functionals $f_s$ and $f_b$ depend on the mean number density $n=(n_{+}+n_{-})/2$ and the charge density $\rho$. These quantities vary only along the $z$ direction. In the bulk, the mean density takes the value $n_0$. In both bulk and near-interface regions, the fields that appear in the free energy are required to satisfy several conditions. The potential should decay to zero in the bulk. The potential must satisfy the Poisson equation for a source equal to the mean-field charge. The electric field at the interface is determined by the external charges. Additionally, the near-interface and bulk regions have equal chemical potentials for the species. All these requirements can be implemented explicitly in the variational functional by means of Lagrange multipliers. For expediency in this presentation, we omit these details.

In the bulk region, the free energy density has the regular solution model form: 
\begin{equation}
f_b=k_B T \sum_j[n_j \ln (v n_j)-1],
\end{equation}
where $j$ runs over the ion species and solvent, and  $v$ is a thermal volume for the solution. The functional can be evaluated and expanded around a uniform bulk state and can be shown to lead to a Poisson-Boltzmann description of the bulk response to external fields. 

Next, we construct the functional for the near-interface region. Here, ions are assumed to be found in aggregates of size $N_c$ that are considered to act as fundamental particles. The entropy of the aggregates within the solvent is described by a regular solution model. In addition, within the aggregates, the ions interact with their neighbors and produce an effective cohesive energy per particle, $E_c$. This energy arises mainly from electrostatic interactions. We note, however, that the presence of the interface imposes important steric constraints that facilitate the organization of the ionic aggregates. Thus, this average free energy per particle still contains information about the entropy of the aggregate. For simplicity, all ions within the near-interface region are assumed to be present in these aggregates. In the near-interface region, the model explores the possibility of a number density $n_w$, with a step function profile uniform through the region, that might be different from the bulk value $n_0$. The associated free energy density is 
\begin{equation}\label{eq:fc}
f_s=2nE_c+\sum_j{n_j}[\ln(v n_j)-1]. 
\end{equation}
The second term in Equation \ref{eq:fc} is the entropy of a mixture of only two species: the aggregates with number density $n_c=2n_w/N_c$, and the solvent. 

It is important to emphasize that the cohesive energy $E_c$, though mostly of electrostatic origin is not captured by the mean field contribution proportional to $\rho\phi$.  This is the case, for example, in ionic crystalline solids where, when averaged over large length scales, both charge density and field are zero. Instead, the cohesive energy is well described by a Madelung constant, which evaluates the sum of Coulomb interactions of the whole system with a representative member of the lattice. This idea has been used to model macroions collapsed with their counterions and to describe the adsorption of ions to charged walls \cite{rouzina1996macroion,shklovskii1999wigner,solis2000collapse}. The precise value of the cohesive energy is difficult to calculate but it should be nearly independent of the bulk concentration. Thus, we use the assumption that $E_c$ is a constant and set its value to obtain agreement with observed simulation results. It is important to note, however, that we assume that this term is only present near the interface. 

Once the functional is specified, standard variational arguments indicate that the best approximation to the properties of the system is obtained when a selection of ion number density and charge distribution minimizes the free energy. For this, it is necessary to explore all possible ion distributions. However, these configurations are limited by steric effects. 
To make calculations feasible, as well as to incorporate features of the observed behavior in simulations, we consider only step-wise charge  distributions, with steps corresponding to  positions $(k-1/2)d$ away from the interface, where $k$ is an integer. That is, we assume that within the aggregates, there are well-defined layers of ions. This is consistent with simulation results (Figures \ref{fig:rho_nacl}, \ref{fig:netrho_nacl}, \ref{fig:sf_0474lowchargesurface}, \ref{fig:snapshots}) showing enhanced occupation of these near-interface regions by ions. The charge density $\rho$ can only take values from $-2e n_w$ to $+2e n_w$ and can be expressed as a fraction of the mean ion number density: $\rho=p_k 2 e n_w$, where $-1\leq p_k\leq1$ and $e$ is the elementary charge. $p_k$ refers to the charge fraction associated with the layer number $k$.
Thus, the possible charge distributions considered are
\begin{equation}
\rho(z)=2 e p_k  n_w, \,\,\,  (k-1)d < z \leq kd, \,\, \mathrm{for} \,\, k = 1, \ldots, L.
\end{equation}  

For a given bulk particle density, and a prescribed surface charge density, the free energy functional can now be minimized with respect to the charge fraction $p_k$ in the layers, the number density $n_w$ near the interface, and the total number of structured  layers $L$. This minimization is carried out under the constraints noted above. Finally, while the number of layers $L$ is variable in principle, we only consider cases where these layers exhibit non-negligible charge density. We discard possibilities that extend the region of aggregates  indefinitely into the bulk. This is in accordance with the observation that the external planar interface has an ordering effect that is not likely to extend into the bulk. 
 
\begin{figure}[ht]
	\centering
	\includegraphics[width=0.48\textwidth]{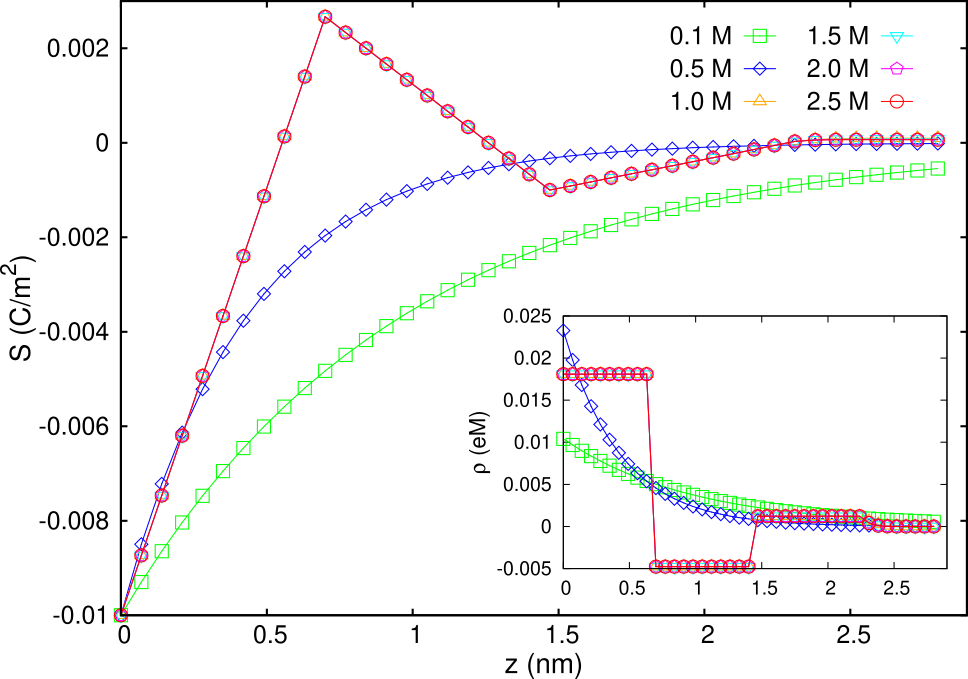}
	\caption{Results from the phenomenological model for the screening factor of a size-symmetric monovalent electrolyte near an interface characterized with surface charge density $-0.01 \mathrm{C}/\mathrm{m}^2$. Different symbols represent different salt concentrations $c \in 0.1 - 2.5$ M. Inset shows the charge densities.}
	\label{fig:theory}
\end{figure}

Analysis of the model produces the following results. For a given value of the surface charge density, as the bulk concentration increases, a transition is observed  from a fluctuation-dominated state without a distinct near-interface behavior, to a state exhibiting structured layers of finite thickness near the interface. A concrete example is shown in Figure \ref{fig:theory} where the screening factors and the associated charge densities are provided. 
The parameters used roughly correspond to simulation conditions for size-symmetric electrolyte ions with diameter $d$ equal to the Bjerrum length $l_B$ in water, and $\sigma_s = -0.01 \mathrm{C}/\mathrm{m}^2$. The example assumes an aggregate size of $N_c=30$, but the results are insensitive to this selection and are nearly identical for values above this size.

The onset of aggregate formation corresponds to concentrations of 1 M. To observe this transition the required cohesive energy is $E_c=-0.8 k_BT$. We find that the transition is sharp and a near-interface region of structured layers of thickness $L=3$ is preferred above the concentration threshold (crossover concentration). In Figure \ref{fig:theory} this region extends up to $z \approx 2.1$ nm. Within the near-interface region, the ionic density is higher than in the bulk, in contrast with the Poisson-Boltzmann case, where it remains the same and increases slowly with bulk concentration. Above the transition, the charge density is effectively independent of the bulk concentration. This density, in all cases, is oscillating. The first layer reverses the charge of the interface, the second layer of charge again inverts the net charge and so does the third. This oscillatory nature of the charge density makes the screening factor non-monotonic. 
As in this regime the  charge density is effectively independent of the bulk concentration, the screening length becomes a constant: $z^* = L d$,  i.e., three ion diameters in the model considered.  
We emphasize that this behavior is in sharp contrast with the low concentration regime where the screening factor has a monotonic behavior, and the resulting screening length is shortened with increasing concentration. 

\section{Discussion and Conclusion}

Using molecular dynamics simulations of the primitive model of electrolytes, we performed a systematic study of the ionic structure of aqueous monovalent electrolyte solutions confined by two planar interfaces over a wide range of electrolyte concentrations $c \in (0.1, 2.5)$ M, interfacial separations $h \in (5,8)$ nm, surface charge densities $\sigma_s \in (-0.005,-0.02)$ C/m$^2$, and counterion sizes $d_+ \in (0.2 - 0.63)$ nm.
Our focus was on understanding the behavior of ions in nanoconfinement created by interfaces under high electrolyte concentrations and how the ionic structure influences the screening of the charged interfaces. 
The ionic structure was quantified by evaluating the density profiles of ions, net charge densities, screening factors, and decay lengths associated with the screening of the charged interface. 

Results show the presence of two distinct regimes of screening behavior as the concentration is changed from 0.1 M to 2.5 M for electrolytes with cations and anions of sizes corresponding to hydrated sodium and chloride ions. For low $c \lesssim 1$ M, the screening factor exhibits a monotonic decay to 0 with a decay length that decreases sharply with increasing $c$. On the other hand, for high $c \gtrsim 1$ M, the screening factor has a non-monotonic, oscillatory behavior signaling charge inversion and formation of structured layers near the interfaces. The decay length under these conditions rises with increasing $c$.

The changes in the screening behavior for electrolytes are observed over a wide range of systems generated by tuning the interfacial separation, surface charge density, and the size of the ions. 
The distinct regimes of the screening behavior are attributed to the dramatic changes in the ionic structure with increase in $c$ including the enhanced accumulation of both counterions and co-ions near the interface and the non-monotonic behavior of the net charge density. Both these changes are driven by the rise in the strength of the steric ion-ion correlations. When the ion size is reduced to $d_+ = d_- = 0.313$ nm, the screening behavior for electrolytes within the concentration range $c \in (0.1, 2.5)$ M does not exhibit any significant deviations from the Debye-Huckel scaling of the decay length as a function of concentration.
 
Our results directly probe the effect of increasing the electrolyte concentration on the screening of the charged surfaces. Recent studies have examined these effects by extracting the correlation length associated with the charge-charge pair correlation functions in bulk electrolyte solutions \cite{coles2020correlation}. 
These studies have found that the correlation length rises with concentration for sufficiently high $c$ as a power law. The exact scaling relation between the correlation length and concentration remains unclear; different approaches including liquid-state theories, atomistic simulations, and experimental studies place the power-law exponent $n$ to be between 1 and 3 \cite{coles2020correlation,perez2017underscreening}. Our finding of $n \approx 1.5$ derived by examining the decay of the screening factor near a charged surface screened by electrolyte ions is within this range. 

We realize that, although our results are accurate for the primitive electrolyte model, the model system itself assumes a homogeneous structureless solvent. 
Work on quantifying the effects of solvent structure is under way, and initial results of atomistic simulations of confined aqueous NaCl solutions show a screening behavior qualitatively similar to that obtained with the implicit-solvent model including the transition of the screening factor from monotonic decay to non-monotonic decay as $c$ is increased. 
A comprehensive study based on simulations of explicit-solvent models is needed to distill the contributions of solvent effects vs. ionic correlations toward the evolution of structural features and associated changes in the screening behavior.

We note that theoretical studies using both implicit-solvent and explicit-solvent descriptions have predicted the rise in the decay length with increasing $c$ for high $c$ conditions \cite{attard1993asymptotic,coles2020correlation}. However, the two models can yield distinct quantitative scaling behavior, e.g., the power-law exponent associated with the rise in the decay length can be different for a two-component model (cations and anions) compared to a three-component model (cations, anions, solvent particles) \cite{coles2020correlation}.

Our model system assumed unpolarizable material surfaces. The dielectric permittivity of the solvent was also assumed to be the same for all electrolyte concentrations. In our earlier work, we found the effects of polarization charges to be generally weaker for monovalent electrolytes at concentrations $\gtrsim 0.1$ M \cite{jing2015ionic}. Further, we performed simulations of monovalent electrolyte ions in water (dielectric permittivity $\epsilon = 80$) confined between two uncharged, polarizable surfaces separated by 5 nm and characterized with material permittivity $\epsilon_m = 2$ using methods outlined in previous papers \cite{jing2015ionic,jadhao2012simulation,jadhao2013variational}. 
These simulations showed that surface polarization charges change the ionic structure minimally for 0.5 M and the effects are much more suppressed for 2.0 M. The effects of polarization charges are expected to be further overwhelmed by the free surface charges in the presence of charged interfaces.
Furthermore, we note that our results for the primitive model system can offer pathways to quantify those additional effects arising due to polarizable surfaces. 

Finally, we note that the deviation in the scaling behavior of decay length vs. concentration can occur even at lower concentrations when electrostatic coupling is higher (e.g., with multivalent ions or under conditions of low temperature or solvents with low dielectric permittivity) \cite{kjellander2019intimate}. Our results do not describe these effects. 

To complement the MD simulation results, we constructed a minimal phenomenological model for the system that explores a subset of possible ion number and charge densities.  The model can be considered as a simplified version of the more detailed density functional approaches \cite{de2020interfacial,ma2020classical} that  nevertheless highlights the key features of the system.
Two features not included in the model are the smooth decay of the number density into bulk and the detailed structure of the species within the near-interface region. We have not explicitly modeled the steric effects of the interface and their decay into the bulk that would lead to smoother behavior of the number density. Description of the distributions of ions of different species is also not directly considered, which limits our results to the size-symmetric electrolyte. 

The model implements the observation that, near the charged interface, both simulations and the simplest interpretations of experimental results \cite{smith2016electrostatic}, indicate the presence of strongly correlated structures. Beyond articulating this observation, the model ties together other aspects of the system behaviour. The structured region near the surface is described by a free energy functional that scales differently than at the bulk. As a result, we obtain number density profiles higher near the interface. This is consistent with observations of increased ion density in Figures \ref{fig:rho_nacl}(d-f) and \ref{fig:rho_highsigma}(c-d), which are not found at lower concentration and are not predicted by Poisson-Boltzmann descriptions of charge accumulation at charged surfaces. For high concentrations, when the structured state of the model is preferred, the system can still select among charge distributions that are consistent with a layered structure. As Figure \ref{fig:theory} shows, this density is always oscillating and, within the model largely independent of the bulk concentration, leading to constant decay lengths in this regime. The simulations show smoothly changing charge densities, that nevertheless preserve peak locations consistent with layering, and decay lengths that increase slightly at high concentrations. 

\begin{acknowledgments}
This research was supported by the National Science
Foundation through Awards 1720625 and DMR-1753182. Simulations were performed using the Big Red II and III supercomputing systems. 
\end{acknowledgments}

\appendix 

\section{Supplementary Results}

\begin{figure}[ht]
\centering
\includegraphics[width=0.48\textwidth]{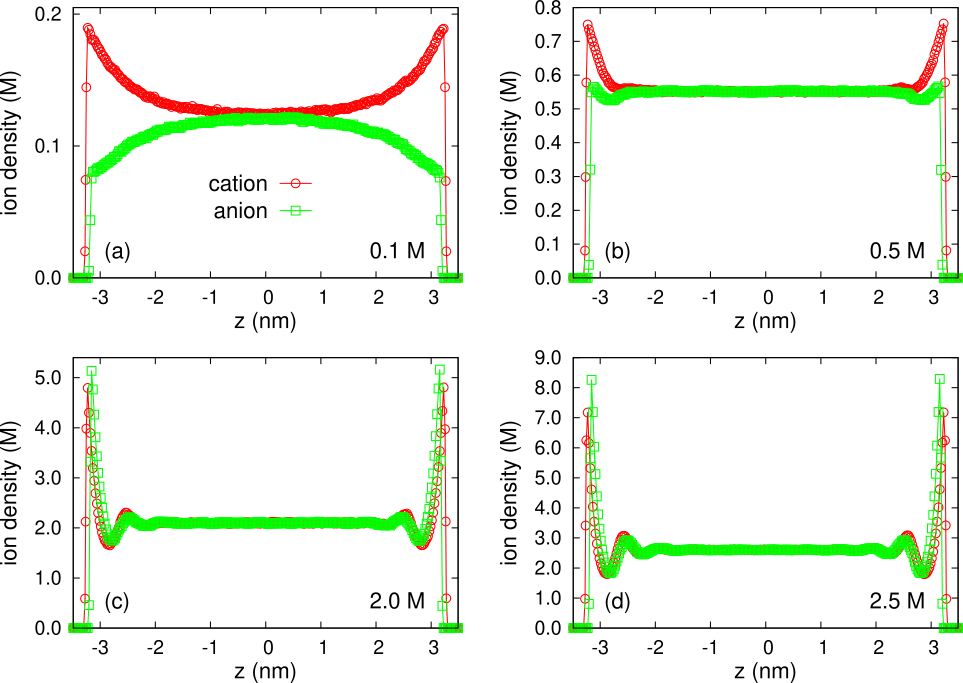}
\caption{
Density profiles of cations (circles) and anions (squares) for the electrolyte system discussed in Section \ref{sec:rho} when the ions are confined between two interfaces separated by $h=7$ nm. Results are shown for 0.1 M (a), 0.5 M (b), 2.0 M (c) and 2.5 M (d).
}
\label{fig:rhoat7nm}
\end{figure}

Figure \ref{fig:rhoat7nm} shows the ionic density profiles for the electrolyte system discussed in Section \ref{sec:rho} when the ions are confined between two negatively-charged interfaces separated by $h=7$ nm at 0.1 M (a), 0.5 M (b), 2.0 M (c) and 2.5 M (d). Results are similar to the density profiles shown in Figure \ref{fig:rho_nacl} for the case where the interfacial separation is 5 nm. At 0.1 M, cations accumulate near the interfaces while anions are depleted near the interfaces. However, for concentration $c\gtrsim 2$ M, both cations and anions exhibit enhanced accumulation near the interfaces.

\begin{figure}[ht] 
\centering
\includegraphics[width=0.48\textwidth]{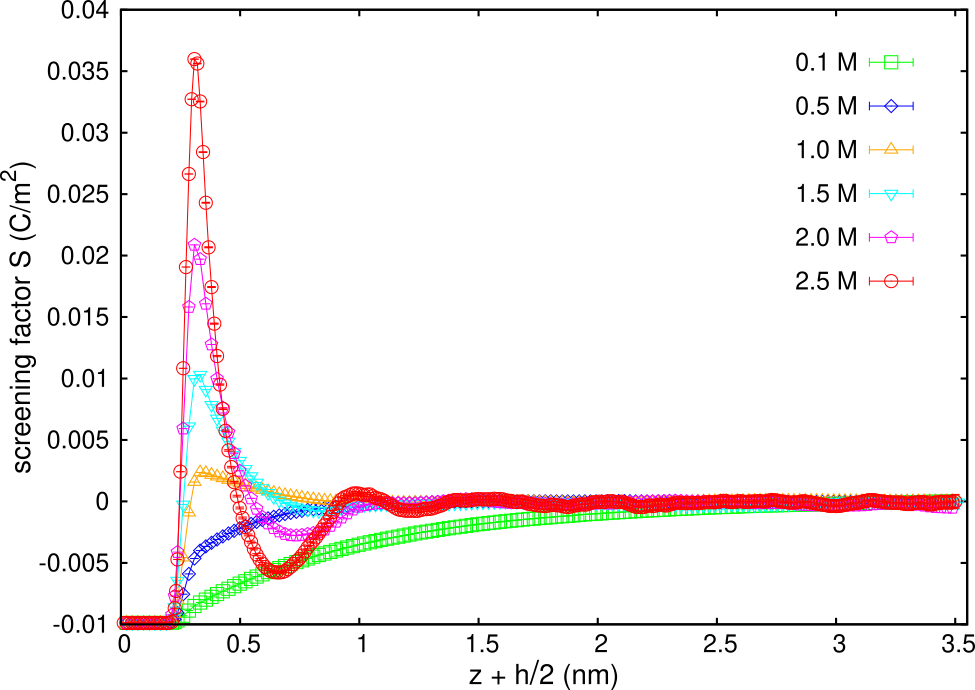}
\caption{Screening factor vs. the distance from the left interface for the electrolyte systems considered in Figure \ref{fig:sf_0474lowchargesurface} when ions are confined within an interfacial separation of $h=7$ nm at different electrolyte concentration $c \in (0.1, 2.5)$ M.}
\label{fig:sf_7nm}
\end{figure}

Figure \ref{fig:sf_7nm} shows the result for screening factor $S(z)$ vs. $z + h/2$ for the electrolyte system considered in Section \ref{sec:sf} confined between two charged interfaces with separation of $h=7$ nm. Results are similar to the screening factor $S$ for $h=5$ nm (Figure \ref{fig:sf_0474lowchargesurface}). 

\begin{figure}[ht]
\centering
\includegraphics[width=0.48\textwidth]{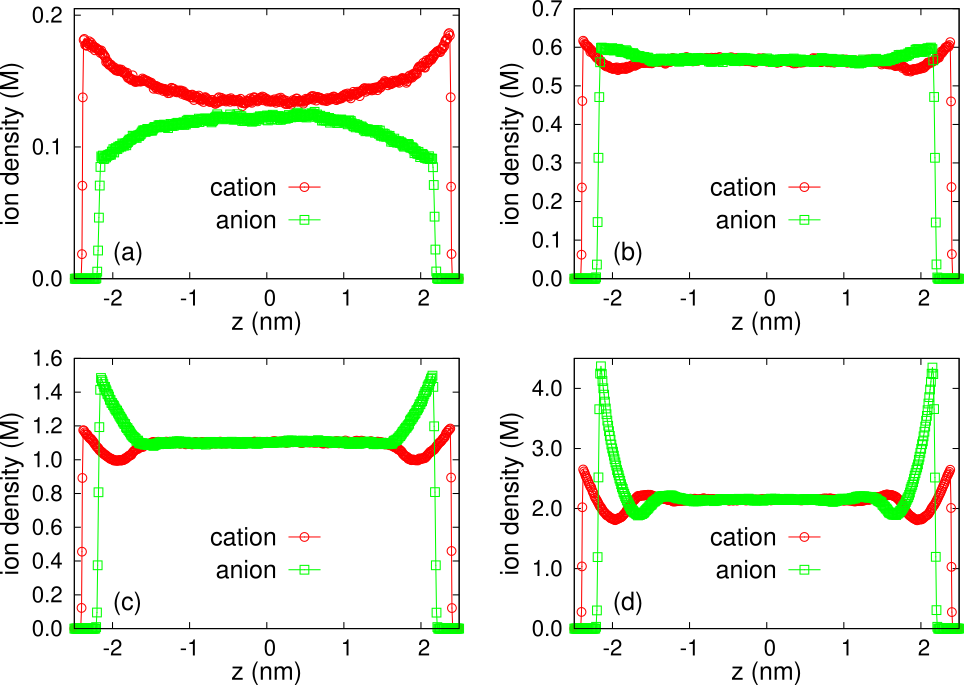}
\caption{Density profiles of cations (circles) of diameter 0.209 nm and anions (squares) of diameter 0.627 nm confined within two interfaces separated by 5 nm and characterized with a surface charge density $-0.01 \mathrm{C}/\mathrm{m}^2$. Results are shown for 0.1 M (a), 0.5 M (b), 1.0 M (c), and 2.0 M (d).}
\label{fig:rho209}
\end{figure}

\begin{figure}[ht]
\centering
\includegraphics[width=0.48\textwidth]{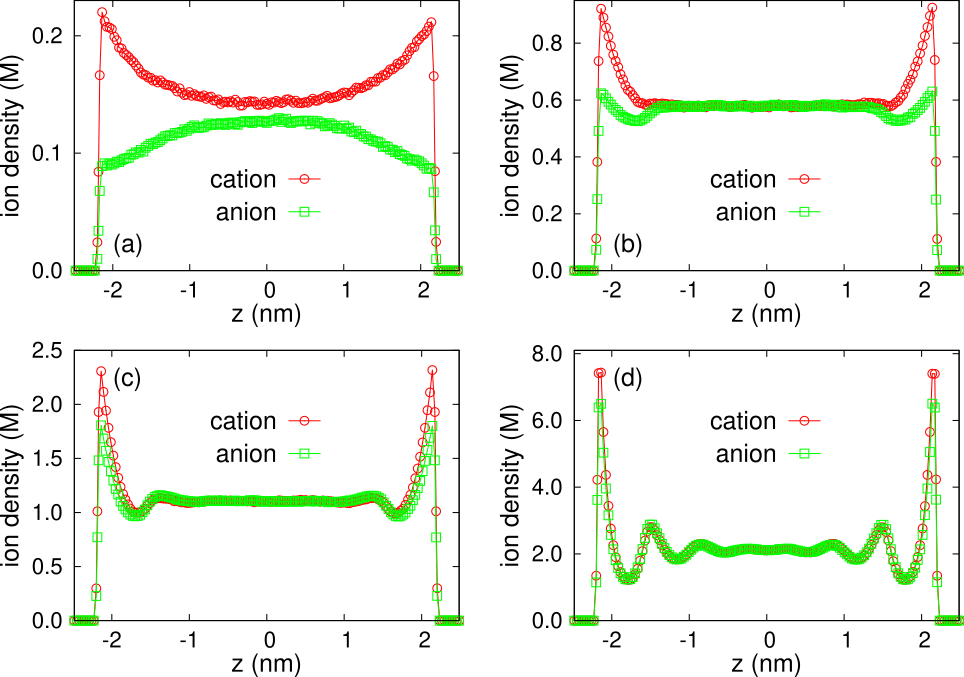}
\caption{Density profiles of cations (circles) of diameter 0.627 nm and anions (squares) of diameter 0.627 nm. Other parameters are the same as in Figure \ref{fig:rho209}. Results are shown for 0.1 M (a), 0.5 M (b), 1.0 M (c), and 2.0 M (d).}
\label{fig:rho627}
\end{figure}

Figures \ref{fig:rho209} and \ref{fig:rho627} show the ionic density profiles corresponding to the electrolyte system for which the net charge density profiles were shown in Figure \ref{fig:netrho_ionsize} in the main text. Figure \ref{fig:rho209} shows the ionic distributions for the system with cations of diameter $d_+= 0.209$ nm and anions of diameter $d_- = 0.627$ nm. Figure \ref{fig:rho627} shows the results for the same system when cations of diameter $d_+= 0.627$ nm are considered.
The anion peak density for $c \gtrsim 1.0$ M for the system with cations of size $0.209$ nm is higher than the cation peak density near the interface. On the other hand, under similar conditions, the anion peak density near the interface is smaller compared to the cation peak density for the size-symmetric electrolyte system with cations of size $0.627$ nm.

\section*{Data Availability}
The data that support the findings of this study are available within the article or from the corresponding author upon reasonable request.

\end{document}